\documentclass[amsmath,amssymb,prx,twocolumn,showkeys,showpacs,superscriptaddress,floatfix,longbibliography]{revtex4-2}

\usepackage{graphicx}
\usepackage{dcolumn}
\usepackage{bm}
\usepackage{booktabs}
\usepackage{braket}
\usepackage{graphicx}%
\usepackage{multirow}%
\usepackage{amsthm}%
\usepackage{mathrsfs}%
\usepackage[title]{appendix}%
\usepackage{xcolor}%
\usepackage{textcomp}%
\usepackage{booktabs}%
\usepackage{algorithm}%
\usepackage{algorithmicx}%
\usepackage{algpseudocode}%
\usepackage{listings}%
\usepackage{dsfont}
\usepackage{hyperref}
\usepackage[normalem]{ulem}
\usepackage{epstopdf}
\usepackage{tabularx}

\begin{document}


\title{Single-step SU(3) Hahn Echo in a Qutrit Atom Interferometer}

\author{Chirantan~Mitra}
 \affiliation{School of Physical and Mathematical Sciences, Nanyang Technological University, 21 Nanyang Link, Singapore 637371, Singapore}
\author{Lucas~Gabardos}%
  \affiliation{School of Physical and Mathematical Sciences, Nanyang Technological University, 21 Nanyang Link, Singapore 637371, Singapore}
\author{François~Impens}
\affiliation{Instituto de F\'{\i}sica, Universidade Federal do Rio de Janeiro, 21941-972, Rio de Janeiro, Rio de Janeiro, Brazil}%
\author{David~Gu\'ery-Odelin}
\affiliation{Laboratoire Collisions Agr\'egats R\'eactivit\'e, UMR 5589,
FERMI, Universit\'e de Toulouse, CNRS, 118 Route de Narbonne, 31062 Toulouse CEDEX 09, France}
\author{David~Wilkowski}
 \affiliation{School of Physical and Mathematical Sciences, Nanyang Technological University, 21 Nanyang Link, Singapore 637371, Singapore}
\affiliation{MajuLab, International Joint Research Unit IRL 3654, CNRS, Universit\'e C\^ote d'Azur, Sorbonne Universit\'e, National University of Singapore, Nanyang Technological University, Singapore}
\affiliation{Centre for Quantum Technologies, Nanyang Technological University, 50 Nanyang Avenue, Singapore, 639798, Singapore}
\author{Chang~Chi~Kwong}
\email{changchikwong@ntu.edu.sg}
\affiliation{School of Physical and Mathematical Sciences, Nanyang Technological University, 21 Nanyang Link, Singapore 637371, Singapore}
\affiliation{MajuLab, International Joint Research Unit IRL 3654, CNRS, Universit\'e C\^ote d'Azur, Sorbonne Universit\'e, National University of Singapore, Nanyang Technological University, Singapore}
\affiliation{Centre for Quantum Technologies, Nanyang Technological University, 50 Nanyang Avenue, Singapore, 639798, Singapore}
\date{\today}

\begin{abstract}

The Hahn echo is a well-known dynamical decoupling method that refocuses the dephasing due to energy inhomogeneities of a spin-1/2 ensemble and is widely used to extend phase coherence in quantum computers and sensors. Extensions of such techniques beyond the spin-1/2 have been reported, but have mainly used SU(2) gates, limiting their scalability. Here, we generalize the Hahn echo to qutrits using single-step SU(3) gates and extend the coherence time by more than an order of magnitude. The gates are generated through optimal control theory and implemented in an ultracold atomic ensemble in free fall. A quantum process tomography procedure benchmarks the gates and yields fidelities of  $\sim93\%$, limited mainly by state preparation and measurement errors. Due to the geometry of the laser beams, the generalized Hahn echo also realizes a three-path Mach-Zehnder atomic interferometer with two independent interferometric phase differences. Our results therefore establish SU(3) control for dynamical decoupling of qutrits and multi-path atom interferometers, with the latter opening the possibility for vector force measurements and interferometric tests of the validity of Born’s rule.
\end{abstract}

\keywords{qutrit, Hahn echo, quantum information, quantum sensing, ultracold gas}

\maketitle

\section{Introduction}\label{sec1}

In a Hahn or spin echo, an intermediate Pauli-X gate is inserted into a Ramsey sequence to cancel phase errors accumulated due to quasi-static inhomogeneous fields~\cite{Hahn_echo}. This technique was originally used in nuclear magnetic resonance to improve the spin dephasing time~\cite{CPMG_DD}, but it was soon extended to other domains such as neutron scattering and photonics, where it has, for instance, been employed as a highly sensitive method to, respectively, probe the structure of matter \cite{WANG2024100128} and characterize its nonlinear optical response \cite{PhysRevResearch.2.012026}.

The Hahn echo belongs to a larger group of dynamical decoupling techniques which have recently regained significant attention in quantum computing as error‑mitigation tools toward fault‑tolerant computation. They are commonly used to suppress dephasing in qubits due to time-dependent inhomogeneous environmental couplings~\cite{viola_dynamical_1999,viola_random_2005,viola_robust_2003,PhysRevLett.106.240501,PhysRevApplied.20.064027}. In addition to the Hahn echo, they include multi-pulse sequences such as Carr–Purcell–Meiboom–Gill (CPMG)~\cite{CPMG_DD}, XY-4~\cite{gullion_new_1990} or Uhrig dynamical decoupling (UDD)~\cite{Uhrig_DD}, which further suppress other types of environmental noise by averaging out the qubit-environment interactions. Such schemes have been shown to significantly improve the coherence time of qubits during logical operations \cite{han_protecting_2025}, idle times \cite{coote_resource-efficient_2025}, and transportation~\cite{OptConveyorBelt}. The use of dynamical decoupling schemes also extends beyond quantum computing. They have been used, for example, with polar molecules~\cite{Li2023} and magnetic atoms~\cite{Laupretre2026} to study itinerant spin dynamics or with diamond NV centers for quantum metrology~\cite{Zhou2020}.

Recently, Hahn echo and related dynamical decoupling schemes have been extended beyond qubits to $d>2$ dimensional systems or qudits. As they explore a larger Hilbert space, qudits are expected to reduce the circuit depth and the gate count of quantum processors \cite{wang2020qudits,nikolaeva2024efficient}, as well as improve quantum error correction codes~\cite{Campbell2014,Brock2025}. However, an enlarged state space generally introduces additional decoherence channels and increases sensitivity to noise \cite{goss_extending_2023}, requiring more efficient dynamical decoupling protocols. Qudit systems have already been realized in superconducting circuits~\cite{liu_performing_2023}, trapped ions~\cite{Ringbauer}, NV centers \cite{fu_experimental_2022, zhou_robust_2024} and neutral atoms \cite{Lindon_2023, Ammenwerth, Ahmed_2025, Eloi25}. In addition, qudit dynamical decoupling has been proposed~\cite{vitanov_dynamical_2015} and demonstrated~\cite{yuan_preserving_2022, iiyama_qudit-generalization_2024, tripathi_qudit_2025, rubin-osanz_implementing_2026} on sequences constructed from pairwise SU(2) operations. In this case, the number of gates required increases as $\mathcal{O}(d^2)$~\cite{vitanov}. For instance, the echo operation in a qudit Hahn echo requires $d-1$ applications of the cyclic $\hat X_d$ operator, which is itself decomposed into $d-1$ two-level operations. The echo then necessitates $(d-1)^2$ operations overall. Therefore, the accumulation of gate errors may become a bottleneck as the Hilbert-space dimension increases~\cite{kononenko, Ringbauer}.

In this work, we demonstrate an extension of the coherence time of qutrits ($d=3$) by more than an order of magnitude using a generalized Hahn-echo protocol based on single-step SU(3) operations (see Fig.~\ref{fig:1}). These operations are implemented directly as SU(3) transformations, rather than being decomposed into multiple SU(2) rotations. {While there have been recent demonstrations of single-step SU(3) gates \cite{Yu2026, Cho2024, Eloi25}, here we implement such operations using internal atomic transitions and apply them to realize the generalized Hahn echo. We encode our qutrit in three Zeeman ground states of a strontium-$87$ ($^{87}$Sr) ultracold atomic gas, and use optical transitions in a tripod scheme~\cite{PhysRevLett.129.130402} to implement the gate operations.}  
{Since each ground state of the tripod scheme is associated with a momentum state, the main source of dephasing is the finite momentum dispersion of the ultracold gas, which limits the overlap time of the atomic wave packets unless a Hahn echo refocuses them.} 
The gates used for quantum superposition and echo are the Hadamard ($\hat{H}_3$) and generalized Pauli-X ($\hat{X}_3$) gates, respectively. These gates are synthesized with optimal control theory (OCT) \cite{ansel_introduction_2024} applied directly to unitary operations \cite{Eloi25, 4xcd-wyxx, Cho2024, PRXQuantum.4.040333}. OCT is well suited for gate optimization as it allows the derivation of robust drives with respect to environmental noise~\cite{Yang2022,Oda2023,Zhang2023,Shao2024,Baker2025}, hardware constraints~\cite{Machnes2018,Lewis2026,Hu2026} or losses due to spontaneous emission~\cite{Palao2008}. The gates are independently benchmarked via quantum process tomography, yielding gate fidelities of 0.936(11) and 0.930(14) for $\hat{H}_3$ and $\hat{X}_3$ gates, respectively. We identify the state preparation and measurement (SPAM) error as the main source of infidelities.

\section{Results}
\subsection{Hahn-echo for qutrits}\label{sec2}

\begin{figure*}[t!]
 \centering
 \includegraphics[width=0.9\textwidth]{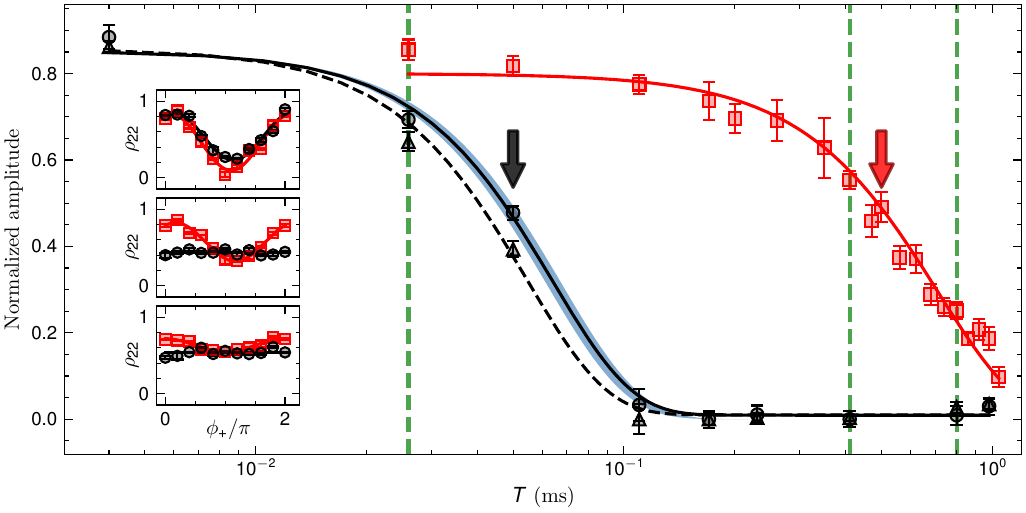}
 \caption{Qutrit dephasing time for generalized Ramsey and Hahn-echo sequences. The black circles (triangles) correspond to the normalized interference amplitude for a generalized Ramsey sequence, obtained by integrating the {state $\ket{2}$ population time-of-flight} signal {over a momentum range of} $\hbar\kappa_r[-0.125,0.125]$ ($\hbar\kappa_r[-0.5,0.5]$), measured as a function of the separation time $T$ between the two Hadamard gates. From Gaussian fits, the Ramsey dephasing times, defined by the $1/e^2$ decay time, are extracted as $T_2^*=91(4)\,\mu\text{s}$ ($77(2)\,\mu\text{s}$). The blue shaded region overlaid with the black circles represents the predicted range of Ramsey profiles for the corresponding momentum integration range, taking into account the experimental temperature and initial horizontal cloud size. The red squares correspond to normalized amplitude decay for the Hahn-echo interferometer. A Gaussian fit to the data yields a lower bound to the $T_2$ homogeneous dephasing time of $1007(21)\,\mu\text{s}$, which is insensitive to the momentum integration range (see Appendix~\ref{secAxx}). The three insets show the interference fringes for the Ramsey sequence (black circles) and Hahn echo (red squares) as a function of the scanned phase $\phi_+$ of the $\sigma_+$ beam, at dark times indicated by the vertical green dashed lines in the main panel. The solid lines in the insets are sinusoidal fits. The black and red vertical arrows indicate data points at normalized amplitudes of $\sim0.5$ for the Ramsey and Hahn-echo interferometers, respectively. The corresponding raw images are shown in Fig.~\ref{fig:2}(e)~and~(f).}
 
\label{fig:1}
\end{figure*}

For qubits, the dephasing time $T_2^*$ is measured by Ramsey interferometry, which consists of two $\pi/2$-pulses separated by a dark time $T$. The inhomogeneous dephasing, accumulated during the dark time, can be compensated for by inserting a Pauli-X gate at time $T/2$, realizing a Hahn echo that brings the qubits back in phase for the second $\pi/2$-pulse and allows the measurement of the homogeneous dephasing time $T_2$. A generalization of the qubit Hahn echo scenario to qudits consists of a $d$-arm interferometer. The state is initially prepared in one of the $\{\ket{n}\}$ energy eigenstates, with $n\in\set{0,1,..,d-1}$; the $\pi/2$-pulses are generalized by Hadamard gates defined as
\begin{align}
\label{eqn:1}
\begin{split}
\hat{H}_d &= \frac{1}{\sqrt{d}}\sum_{n,m=0}^{d-1}(z_d)^{nm}\ket{n}\bra{m},
\end{split}
\end{align}
where $z_d=\exp(-i\frac{2\pi}{d})$ is the $d^{\mathrm{th}}$ root of unity; and the echo now consists of $d-1$ successive applications of the generalized Pauli-X gate 
\begin{align}
\label{eqn:2}
\begin{split}
\hat{X}_d&=\sum_{n=0}^{d-1}\ket{n+1\,\textrm{mod}\, d}\bra{n},
\end{split}
\end{align}
that cyclically permutes the populations and phases among the $d$ states. Both gates represent SU($d$) operations, with $\hat{H}_d$ belonging to SU($d$) up to an irrelevant global phase. To ensure that each arm accumulates the same dynamical phase throughout the sequence, all $d$ dark-time subsequences should be of same duration, namely $T/d$. 

\begin{figure*}[t!]
 \centering
 \includegraphics[width=1\textwidth]{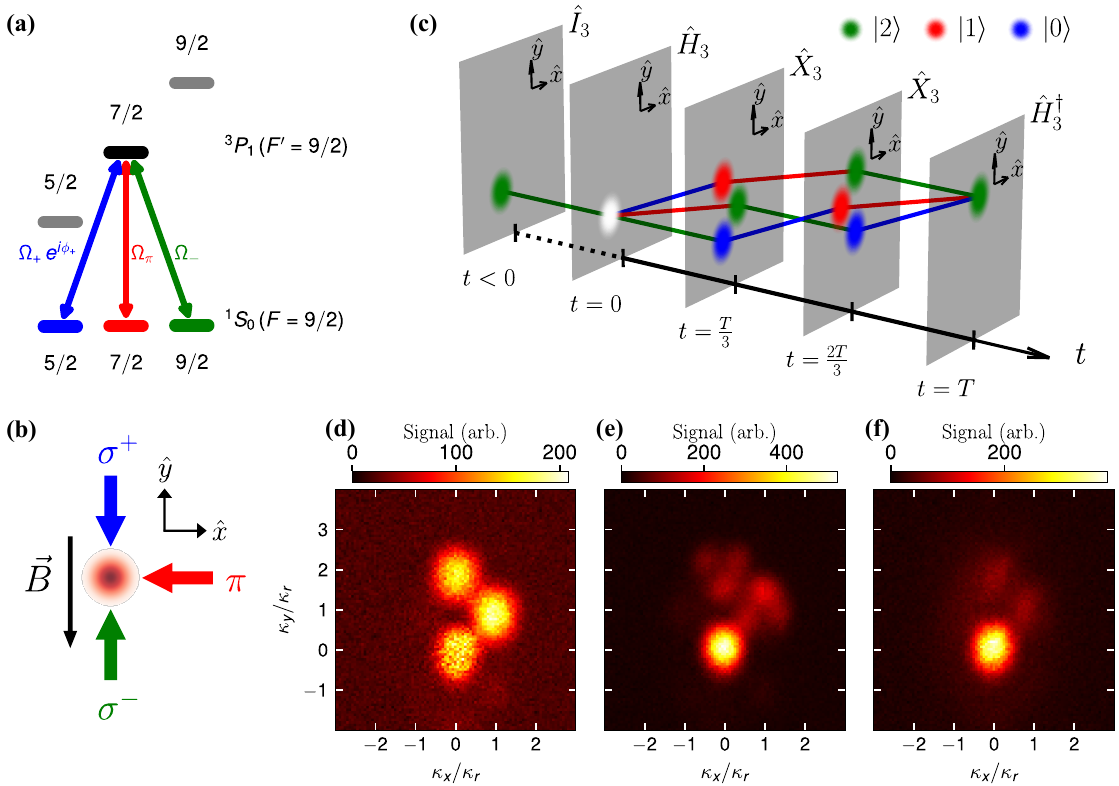}
\caption{\textbf{(a)} Tripod scheme coupling three ground states of $^{87}$Sr to a common excited state. The latter is isolated from the excited manifold by a $B=67.0(3)\,$G bias field. We label the ground states $\ket{0}\equiv\ket{F=9/2,m_F=5/2}$, $\ket{1}\equiv\ket{F=9/2,m_F=7/2}$, and $\ket{2}\equiv\ket{F=9/2,m_F=9/2}$, and associate the colors blue, red, green with these states, respectively. \textbf{(b)} The three optical fields are arranged in the horizontal plane such that a ground-state population transfer is accompanied by a momentum transfer. \textbf{(c)} The qutrit echo sequence starts with the atomic population initialized in state $\ket{2}$. At time t=0, an $\hat H_3$ gate creates a coherent superposition of the three basis states, represented by the white overlap of the green, red, and blue components. The three states then evolve along different spatial trajectories represented by the colored solid lines. At time $t=\frac{T}{3}$, the first $\hat X_3$ gate is applied that cyclically permutes the three states, swapping their momenta in the process. The states continue to evolve along their new trajectories until a second $\hat X_3$ gate at time $t=\frac{2T}{3}$. The interferometer is closed at time $T$ by a $\hat H_3^\dagger$ gate. Without {additional} phase shift, the output state is $\ket{2}$. \textbf{(d)} Time-of-flight image after the first $\hat H_3$ gate. \textbf{(e)} Time-of-flight image after a Ramsey sequence for $\phi_+=0$ and a dark time of $T=0.05\,$ms, corresponding to the black vertical arrow in Fig.~\ref{fig:1}. \textbf{(f)} Time-of-flight image after a Hahn-echo sequence with $\phi_+=0$ and a dark time of $T=0.5\,$ms, corresponding to the red vertical arrow in Fig.~\ref{fig:1}. For (d)-(f) the time-of-flight duration is 15 ms.}

\label{fig:2}
\end{figure*}

The Ramsey and Hahn-echo sequences for $d=3$ are experimentally implemented on a fermionic $^{87}$Sr ultracold atomic ensemble. We use an evaporative cooling technique in a crossed optical dipole trap as the final cooling stage \cite{Hasan_2022}  (see Appendix~\ref{secA2}). We get a degenerate gas of $N=7.8(5)\times10^4$ atoms, with an \textit{rms} width of $8(1)\,\mu$m along the vertical direction and an elliptical density profile in the horizontal plane, characterized by \textit{rms} widths of $9(1)\,\mu$m and $16(2)\,\mu$m along its principal axes. The gas temperature is $0.21(1)T_F$, where $T_F=190(9)\,$nK is the Fermi temperature. About half of the population is in the $\ket{F=9/2,m_F=9/2}$ state of the $^1\!S_0$ ground level, whereas the remaining half is in the $\ket{F=9/2,m_F<0}$ states \cite{PhysRevLett.129.130402}. Then, we switch off the optical dipole trap and let the atomic cloud fall freely. A magnetic field bias of $67.0(3)\,$G is applied to lift the degeneracy of the $^3\!P_1$ excited level (Land\'e factor: $2/33$), whereas the ground level remains quasi-degenerate (Land\'e factor: $-1.3\times 10^{-4}$). As a result, we are able to isolate a quasi-resonant tripod scheme comprising of three ground states $\ket{0}\equiv\ket{F=9/2,m_F=5/2}$, $\ket{1}\equiv\ket{F=9/2,m_F=7/2}$, and $\ket{2}\equiv\ket{F=9/2,m_F=9/2}$, and one excited state $\ket{e}\equiv\ket{F'=9/2,m'_F=7/2}$ (see Fig.~\ref{fig:2}(a)). The ground states of the tripod are used as a qutrit. All tripod transition addressing beams lie in the horizontal plane with directions shown in Fig.~\ref{fig:2}(b). As $\ket{2}$ is the initial state, its position distribution after a time of flight corresponds to a momentum distribution centered at zero; while coherent transfer from $\ket{2}$ to $\ket{1}$ and $\ket{0}$ gives a net momentum change of $\hbar \kappa_r(\mathbf{\hat{x}}+\mathbf{\hat{y}})$ and $2\hbar \kappa_r\mathbf{\hat{y}}$, respectively \cite{PhysRevLett.129.130402}. Here, $\kappa_r=2\pi/\lambda$ is the single-photon-recoil atomic wavenumber, with $\lambda=689\,$nm the $^1\!S_0\,\rightarrow\,^3\!P_1$ intercombination line wavelength (linewidth: $7.5\,$kHz). As the initial momentum dispersion is $0.293(10)\hbar\kappa_r$, smaller than the momentum change due to coherent population redistribution, the three distributions are well resolved after a $15\,\mathrm{ms}$ time of flight (see for example Fig.~\ref{fig:2}(d)). Single-step $\hat{H}_3$ and $\hat{X}_3$ gates, are generated by modulating the phase and amplitude of the tripod beams over $10\,\mu s$. The exact temporal shapes, generated by the OCT, are presented in {Fig.~\ref{fig:3}(a)~and~(b)}.

We measure the dephasing time $T_2^*$, using a qutrit Ramsey interferometer. Starting from the pure state $\ket{2}$, a $\hat{H}_3$ gate is applied before a variable free-evolution time $T$, and finally a $\hat{H}_3^\dagger$ gate closes the interferometer. During the dark time, the phase $\phi_+$ of the $\sigma_+$ beam is shifted to mimic either a phase shift of one arm of the interferometer or a particular displacement of the atomic wave packet due to, for example, an inertial force. The three insets of Fig.~\ref{fig:1} present the resulting interference fringes on the state $\ket{2}$ population at three different dark times (indicated by vertical green-dashed lines). The interference fringe amplitudes (open black triangles) are extracted from sinusoidal fits of the population data and normalized to the expected value at zero temperature. The latter vary from a maximal value of $1$ for $\phi_+=0$ to a minimum of $1/9$ for $\phi_+=\pi$. The Ramsey fringe amplitude decays over a characteristic time corresponding to the $1/e^2$ decay time of its Gaussian fit (black dashed curve). The characteristic time is the dephasing time which is $T_2^*=77(2)\,\mu\textrm{s}$ when the populations are integrated over a momentum window $\hbar\kappa_r[-0.5,0.5]$.

Next, we implement a Hahn echo by applying the $\hat{X}_3$ gate twice during the dark time (see Appendix~\ref{secA1}). We observe an amplitude decay time of $1007(21)\,\mu\textrm{s}$, giving a lower bound to the homogeneous dephasing time $T_2$, see Fig.~\ref{fig:1}. This represents more than an order-of-magnitude improvement over $T_2^*$. 

The main source of inhomogeneity is attributed to the non-zero momentum distribution of the degenerate Fermi gas that limits the spatial extent of the wave packets and thus the overlap of the three qutrit states. This also translates to momentum-induced dynamical dephasing during dark time. The Hahn-echo sequence closes the interferometer, suppressing of the differential phase accumulation along the three paths (Fig.~\ref{fig:2}(c)). 

The inhomogeneous nature of dephasing during the Ramsey sequence results in a spatial structure in the time-of-flight distribution {of the qutrit populations. This is illustrated in Fig.~\ref{fig:2}(e), for a dark time of $T=0.05\,\mathrm{ms}$, where fringes are visible for state $\ket{1}$ and $\ket{0}$, centered at $\boldsymbol{\kappa}_1=(1,1)\kappa_r$ and $\boldsymbol{\kappa}_0=(0,2)\kappa_r$, respectively.} The two-photon detunings of the tripod scheme are set to zero (resonance) for atoms initially in $\ket{2}$ with zero momentum (see Appendix~\ref{secA1}). Therefore, these atoms return to $\ket{2}$ at the end of the Ramsey sequence due to the absence of momentum-induced dephasing. However, this dephasing shows up for non-zero momentum states, leading to final states eventually different from the initial state $\ket{2}$. Importantly, the {Ramsey} dephasing time can be improved by selecting a narrower momentum range in the time-of-flight images. This point is illustrated in the main panel of Fig.~\ref{fig:1}, where restricting the analysis to momentum components in a smaller range $\hbar\kappa_r[-0.125,0.125]$ improves the dephasing time to 91(4)$\mu s$ (see Appendix~\ref{secAxx}). A numerical simulation of the Ramsey sequence, accounting for the momentum integration range and the initial cloud size, yields a dephasing time $T_2^*$ of $91(10)~\mu s$, see blue shaded region in Fig.~\ref{fig:1}, in good agreement with the experiment. Ultimately, the improvement on the dephasing time will be limited by the convolution with initial size of the atomic cloud in the horizontal plane.

In contrast, the time-of-flight images for the Hahn-echo sequence do not show any spatial structures for the populations of state $\ket{1}$ and $\ket{0}$, indicating that the inhomogeneous dephasing due to the momentum dispersion is largely corrected, see Fig.~\ref{fig:2}(f). Finite decay of the Hahn-echo signal is also insensitive to the momentum range, as discussed in Appendix~\ref{secAxx}, and might come from the finite size and residual spatial imperfections of the tripod beams explored during free fall and ballistic expansion of the ultracold gas. 

\subsection{Optimal control design of SU(3) gates}
\label{sec:optimal_control}

We use OCT to engineer quantum gates in the tripod-based optical qutrit platform~\cite{Palao03,Goerz2014}. Our objective is to realize target SU(3) operations within the qutrit subspace, more precisely the $\hat{H}_3$ and $\hat{X}_3$ gates used in the Hahn-echo protocol, while suppressing transient population of the auxiliary excited state $\ket{e}$, associated with the decoherence-sensitive $^3P_1$ level shown in Fig.~\ref{fig:2}(a). To this end, we employ Pontryagin-based OCT~\cite{Boscain21,Eloi25} directly on the unitary propagator $\hat{U}(t)$, neglecting motional dynamics. The optimized pulses implement high-fidelity SU(3) gates in a single step.

The gates are implemented with the tripod scheme shown in Fig.~\ref{fig:2}(a). The experimental controls $\mathbf{u}(t)$ consist of the time-dependent amplitudes $\Omega_{\pm}(t)$ and $\Omega_\pi(t)$ of the $\sigma_\pm$ and $\pi$-polarized tripod beams, together with a time-dependent phase $\phi_+(t)$ applied to the $\sigma_+$ beam; see Fig.~\ref{fig:3}(a)~and~(b). These amplitude and phase modulations are produced using acousto-optic and electro-optic modulators, respectively; see Appendix~\ref{secA3}. The Hamiltonian reads
\begin{eqnarray}
\label{eq:Hu}
\hat{H}(\mathbf{u}(t))& = &  \frac 1 2 \Omega_{+}(t) e^{i \phi_+(t)} \ket{e}\bra{0} + \frac 1 2 \Omega_{\pi}(t) \ket{e}\bra{1}  \nonumber \\
& + & \frac 1 2 \Omega_{-}(t) \ket{e}\bra{2} + \mathrm{H.c.} \, .
\end{eqnarray}

We assume unitary dynamics governed by $\hat{H}(\mathbf{u}(t))$ in the four-dimensional Hilbert space that includes the auxiliary excited state, and denote by $\hat{U}(t)$ the corresponding evolution operator. We define the {four-dimensional unitary OCT fidelity $\mathcal{F}_{\hat{U}}$, used for OCT pulse design,} as
\begin{equation}
\mathcal{F}_{\hat{U}}=
\frac{1}{4^2}
\left|
\mathrm{Tr}\left[
\hat{U}_\mathrm{target,4}^{\dagger}\hat{U}(t_f)
\right]
\right|^2,
\label{eq:definitionFU}
\end{equation}
where $\hat{U}_\mathrm{target,4}=\hat{U}_\mathrm{target}\oplus 1_e$ is the embedding of the $3\times3$ target gate into the full four-dimensional space. Within this unitary description, near-unity fidelity ensures that no residual population remains in the excited state at the final time. We define the cost functional as
\begin{equation}
\label{eq:costfunctional}
J[\mathbf{u}]=\Phi(\hat{U}(t_f))+\int_0^{t_f} dt\, f_0(t),  
\end{equation}
with $\Phi(\hat{U}(t_f))=1-\mathcal{F}_{\hat{U}}$. The running cost
\begin{equation}
f_0(t)=\frac{\lambda}{3}\sum_{j=0}^{2}\left|\Braket{e | \hat{U}(t) | j} \right|^2 \,
\end{equation}
penalizes the transient population of the excited state $\ket{e}$. The parameter $\lambda$ controls the trade-off between final gate fidelity and transient excited-state population, {and is set to $\lambda=0.2/t_f$ for the optimal pulse design. Compared to an optimization based only on gate fidelity $(\lambda=0),$ this value  reduces the time- and initial-state-averaged excited-state population by approximately a factor of four after $10^4$ iterations of the optimization routine,} while maintaining near-perfect final fidelities for the considered gates. 

The optimization is carried out by introducing an adjoint operator $\hat{P}(t)$, of the same dimension as $\hat{U}(t)$, and constructing the associated Pontryagin Hamiltonian (see Appendix~\ref{secA-OCT}). This Hamiltonian generates the coupled evolution of $\hat{U}(t)$ and $\hat{P}(t)$, with the transient excited-state penalty appearing as an inhomogeneous driving term in the adjoint equation. Numerically, the control fields are obtained through an iterative forward--backward procedure. No explicit bounds are imposed on the control parameters. Instead, the total duration $t_f$ is chosen such that the optimized pulse amplitudes remain experimentally realistic, and the resulting OCT pulses are subsequently smoothed using a polynomial fit to remove abrupt variations before experimental implementation.

\subsection{High fidelity Hadamard and Pauli-X gates}\label{sec4}
\begin{figure*}[t!]
 \centering
 \includegraphics[width=1\textwidth]{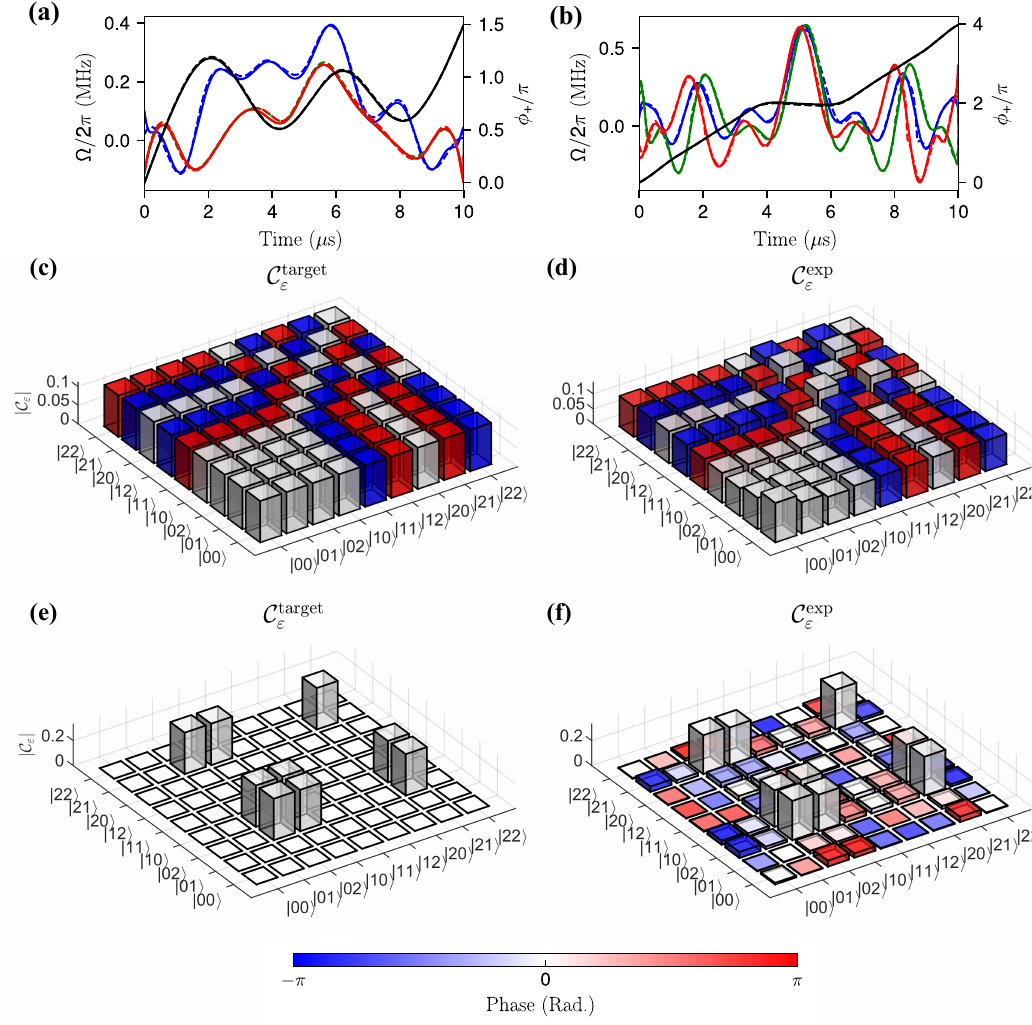}
\caption{\textbf{(a)} OCT-generated pulse sequences for the $\hat H_3$ gate and \textbf{(b)} for the $\hat X_3$ gate. Each sequence consists of three Rabi amplitudes $\Omega_{+}(t)$~(blue), $\Omega_{\pi}(t)$~(red), $\Omega_{-}(t)$~(green) for the $\sigma_+$, $\pi$, and $\sigma_-$ tripod beams, respectively, together with a time-dependent phase $\phi_+(t)$~(black) of the $\sigma_+$ beam. The Rabi amplitudes are referenced to the left y-axis and the $\phi_+(t)$ to the right y-axis. Dashed and solid curves denote the experimentally-generated pulses and their OCT prescription, respectively. \textbf{(c)-(d)} Target and experimentally reconstructed Choi matrices for the $\hat H_3$ gate. \textbf{(e)-(f)} Same as \textbf{(c)-(d)} for the $\hat X_3$ gate.  
}
 \label{fig:3}
\end{figure*}	

A quantum channel is described by a completely positive trace-preserving (CPTP) linear map~\cite{PhysRevA.98.062336} between density operators acting on the input and output Hilbert spaces $\mathscr{H}=\mathbb{C}^d$,
\begin{equation}\label{eqn:8}
\hat \varepsilon:\mathcal{L}(\mathscr{H}_\mathrm{in})\rightarrow\mathcal{L}(\mathscr{H}_\mathrm{out}),
\end{equation}
subject to $\hat\varepsilon \otimes \hat I_d\succeq0$ and $\mathrm{Tr}[\hat \varepsilon(\hat \rho)]=\mathrm{Tr}[\hat \rho]$ for $\hat \rho\in\mathcal{L}(\mathscr{H}_\mathrm{in})$, where $\hat I_d$ is the identity map {and $\mathcal{L}(\mathscr{H}_\mathrm{in})$ is the space of linear operators over the input space}. 
This CPTP map is isomorphic to the density matrix $\mathcal{C}_\varepsilon\in\mathcal{L}(\mathscr{H}_\mathrm{in}\otimes\mathscr{H}_\mathrm{out})$, known as the Choi matrix~\cite{choi_completely_1975, jiang_channel-state_2013}, and defined in the computational basis as 
\begin{equation}\label{eqn:9}
\mathcal{C}_\varepsilon =\frac{1}{d}
\sum_{k,m=0}^{d-1}
\ket{k}\bra{m} \otimes \hat\varepsilon\!\left(\ket{k}\bra{m}\right),
\end{equation}
where $\mathcal{C}_\varepsilon\succeq0$ and $\operatorname{Tr}[\mathcal{C}_\varepsilon]=1$. The map can then be expressed using its Choi matrix as,
\begin{equation}\label{eqn:10}
\hat \varepsilon(\hat \rho) =
d\,\operatorname{Tr}_{\mathscr{H}_\mathrm{in}} \!\left[
(\hat \rho^{T} \otimes \hat I_d) \, \mathcal{C}_\varepsilon
\right],
\end{equation}
where $\operatorname{Tr}_{\mathscr{H}_\mathrm{in}}$ denotes the partial trace over the input system.

The Choi matrices of the $\hat H_3$ and $\hat X_3$ gates are experimentally determined via quantum process tomography (QPT)~\cite{dariano_quantum_2001, poyatos_complete_1997}, by applying the operator to an informationally complete set of input states and making a set of projective measurements on the resulting output states. State preparation and measurement (SPAM) are accomplished by stimulated Raman adiabatic passage (STIRAP) and two-photon Raman rotations. The Choi matrix is obtained through a constrained fit to the experimental data (see Appendix~\ref{secA4}). 

Figure~\ref{fig:3}(a)~and~(b) compares the OCT prescriptions with the experimentally generated pulses for the $\hat H_3$ and $\hat X_3$ gates, respectively. The inverse gate $\hat H_3^\dagger$ used in the interferometer is obtained from the $\hat H_3$ control sequence by reversing the sign of $\phi_+(t)$. The reconstructed Choi matrices of the $\hat H_3$ gate are shown in Fig.~\ref{fig:3}(c)~and~(d), corresponding respectively to the target and experimental cases. The same comparison for the $\hat{X}_3$ gate is presented in Fig.~\ref{fig:3}(e)~and~(f). In both cases, the bare-state populations are obtained after a $15$~ms time of flight and integration over the momentum range $\hbar\kappa_r[-0.125,0.125]$, as for the generalized Ramsey experiment. 


We characterize the performance of the gates using the gate fidelity defined as~\cite{NIELSEN2002249}
\begin{equation}\label{eqn:11}
    \mathcal{F}_a = \frac{d\cdot\mathrm{Tr}[\mathcal{C}^{a\dagger}_\varepsilon\mathcal{C}^\mathrm{target}_\varepsilon] + 1}{d + 1},
\end{equation}
where $a$ stands for OCT, simulated or experimental values. For a theoretical fidelity estimation that is restricted to the computational subspace, the Choi overlap reduces to the qutrit process fidelity
\begin{equation}\label{eq:ideal_Choi_overlap}
\mathrm{Tr}\!\left[\left(\mathcal{C}_\varepsilon^{\overline U}\right)^\dagger
\mathcal{C}^\mathrm{target}_\varepsilon\right]
=\frac{1}{d^2}
\left|
\mathrm{Tr}\!\left[ \hat
U_\mathrm{target}^{\dagger} \overline{U}
\right]
\right|^2.
\end{equation}
$\overline{U}$ denotes the restriction of the full evolution operator $\hat U(t_f)$ to the qutrit subspace.

Using Eqs.~\eqref{eqn:11} and \eqref{eq:ideal_Choi_overlap}, we compute gate fidelities $\mathcal{F}_{\overline U}$ corresponding to the ideal OCT pulses. Those values as well as the ones for simulations and experiments are shown in Table~\ref{tab:gate_fidelities}. The simulations are performed at three successive levels of realism (see Appendix~\ref{secA_simulation}):
\begin{itemize}
    \item idealized simulations using experimentally recorded pulse shapes giving $\mathcal{F}_\mathrm{sim}\sim0.99$;
    \item additionally including initial atomic position and momentum distributions giving $\mathcal{F}_\mathrm{sim}^{\delta p,\delta r}\sim0.98$;
    \item and including the complete tomography sequence with SPAM operations giving $\mathcal{F}_\mathrm{sim}^{\delta p,\delta r + \mathrm{SPAM}}\sim0.95$.
\end{itemize}
The high $\mathcal{F}_\mathrm{sim}$ and $\mathcal{F}_\mathrm{sim}^{\delta p,\delta r}$ fidelities indicate that the pulse profiles are well reproduced in the experiment, and the gates are mostly insensitive to the initial momentum and position distributions. The significant loss of fidelity in $\mathcal{F}_\mathrm{sim}^{\delta p,\delta r + \mathrm{SPAM}}$ shows that SPAM errors are the dominant factor for infidelity, explained by experimental limitations and sensitivity of the Raman rotations to the momentum dispersion {(see Appendix~\ref{secAxx})}.

The experimental gate fidelities $\mathcal{F_\mathrm{exp}}\sim0.93$, calculated from the experimental Choi matrices using Eq.~\eqref{eqn:11}, show slightly lower values than $\mathcal{F}_\mathrm{sim}^{\delta p,\delta r + \mathrm{SPAM}}$, indicating additional sources of infidelity not taken into account by the simulations. Since the OCT pulses act quasi-resonantly on the tripod scheme, additional errors arise from shot-to-shot fluctuations of the bias magnetic field, which has an $rms$ value of $\sim0.1\%$. Numerical estimations give a contribution of the magnetic field fluctuations to gate infidelities at a $\sim 1.2\times10^{-2}$ level. Other noise contributions are smaller than the experimental fidelities' statistical errors. An estimate of the errors is given in Appendix~\ref{secA5}.

\begin{table}[h]
\caption{\label{tab:gate_fidelities}%
$\mathcal{F}_{\overline{U}}$ is the gate fidelity for ideal OCT pulses. $\mathcal{F}_\mathrm{sim}$ is obtained from QPT of numerical simulations using the experimental pulse profiles of the gates; $\mathcal{F}_\mathrm{sim}^{\delta p,\delta r}$ from simulations which account for the initial momentum and position of the atoms and integrate over the momentum range used to extract the experimental populations; and $\mathcal{F}_\mathrm{sim}^{\delta p,\delta r + \mathrm{SPAM}}$ from simulations which include the complete QPT sequence, incorporating SPAM. Error bars come from resampling the initial positions (see Appendix~\ref{secA_simulation}). $\mathcal{F}_\mathrm{exp}$ denotes the gate fidelity obtained from experimental data. All fidelities {are evaluated using} Eq.~\eqref{eqn:11}.}  
\begin{ruledtabular}
\begin{tabular}{c c c c c c}
Gate & {$\mathcal{F}_{\overline{U}}$}
& $\mathcal{F}_\mathrm{sim}^{}$ & $\mathcal{F}_\mathrm{sim}^{\delta p,\delta r}$ & $\mathcal{F}_\mathrm{sim}^{\delta p,\delta r + \mathrm{SPAM}}$ & $\mathcal{F}_\mathrm{exp}^{}$  \\
\midrule
$\hat H_3$ & 0.9997 & 0.9904 & 0.984(3) & 0.958(3) & 0.936(11)    \\
$\hat X_3$ & 0.9990 & 0.9900 & 0.980(2) & 0.952(2) & 0.930(14)    \\ 
\end{tabular}
\end{ruledtabular}
\end{table}

\section{Conclusion and outlook}\label{sec8}

We experimentally demonstrated OCT-engineered single-step SU(3) Hadamard and generalized Pauli-X gates that were implemented in Ramsey and Hahn-echo interferometers using a tripod scheme on an ultracold ensemble of $^{87}$Sr atoms. Our gates were independently benchmarked via quantum process tomography, yielding gate fidelities of $0.936(11)$ and $0.930(14)$ for Hadamard and generalized Pauli-X gates, respectively. Using simulations, we identified SPAM errors as the main source of infidelity. 

These gates were used to implement interferometric sequences on ultracold atom ensembles in free fall. We determined that the interplay between the finite momentum distribution and the momentum exchange during state transfer is the main source of dephasing in the Ramsey configuration. Using the Hahn echo in its generalized SU(3) version, we close the external spacetime paths of the interferometer and homogenize the dynamical phase accumulation among the three paths, leading to an increase of the coherence time by almost a factor of $11$. This work therefore establishes a route towards qudit dynamical decoupling based on single-step SU($d$) operations, avoiding decompositions into sequences of binary SU(2) rotations.

Interestingly, the SU(3) Hahn echo can be viewed as a three-path generalization of the atomic Mach-Zehnder interferometer (MZI). Depending on the geometry, the MZI can be sensitive to an inertial force along a single effective direction \cite{PhysRevLett.67.181} and rotation orthogonal to the laser plane \cite{PhysRevA.80.063604}. Vector acceleration measurements have also been demonstrated by either using sequentially interrogated axes in free space \cite{templier2022tracking} or 2D optical lattice interferometers~\cite{ledesma2025vector}. Here, the non-collinear geometry of the tripod beams allows the three-path interferometer phase to be sensitive to a 2D vector force and, through a double-diffraction geometry~\cite{PhysRevLett.122.043604}, could also be simultaneously sensitive to both a 2D force and rotation. In addition, a genuine three-path interferometer, as presented in this work, could also be used to test the validity of Born's rule~\cite{born_zur_1926} by probing the existence of higher-order interference terms~\cite{covey_probing_2025, valentini_beyond_2022, PhysRevA.101.052111}.

\begin{acknowledgments}
This project is supported by 
the Singapore Ministry of Education Academic Research Fund Tier2 (Grant No. MOET2EP50220-0008), and the Institut Universitaire de France.  F.I. acknowledges support from the Brazilian agencies CNPq (305638/2023-8) and FAPERJ (210.570/2024), and from the CAPES-COFECUB~(20232475706P) program.
\end{acknowledgments}

\section*{Author Contributions}

C.M., C.C.K. and D.W.  conceptualized the experiment. C.M. and L.G. implemented and calibrated the optical control pulses, performed the experiments, collected and analyzed the data. L.G. and C.M. performed the numerical modeling. F.I. and D.G.-O. performed the optimal control design of the gates. C.M. wrote the first draft of the paper. All authors contributed to the writing of the manuscript. D.W. supervised the work.

\section*{COMPETING INTERESTS}
The authors declare no competing interests.

\section*{DATA AVAILABILITY}
The data that support the findings of this study are available from the corresponding author upon reasonable request.

\appendix

\section{Quantum gate optimization with the Pontryagin maximum principle}\label{secA-OCT}

We provide here further details on the optimization procedure, which relies on gradient-based methods derived from the Pontryagin maximum principle~\cite{pontryagin1962,khaneja2005,reich2012,Boscain21}. The state variable corresponds to the four-dimensional unitary evolution operator $\hat{U}(t)$, which satisfies the Schr\"odinger equation
\begin{equation}
\dot{\hat{U}}(t) = -i \hat{H}(\mathbf{u}(t)) \hat{U}(t), \qquad \hat{U}(0)=\mathbb{I}.
\end{equation} 
where the Hamiltonian $\hat{H}(\mathbf{u}(t))$ is given in Eq.~\eqref{eq:Hu}. Introducing an adjoint operator $\hat{P}(t)$, with the same dimension as $\hat{U}(t)$, we define the Pontryagin Hamiltonian
\begin{equation}
\mathcal{H}_P(\hat{U},\hat{P},\mathbf{u},t)
=
-f_0(t)
+\mathrm{Im}\,\mathrm{Tr}\!\left[  \hat{P}^\dagger(t)   \hat{H}(\mathbf{u}(t)) \hat{U}(t) \right],
\end{equation}
which is an optimization Hamiltonian, distinct from the physical Hamiltonian $\hat{H}(\mathbf{u}(t))$.
The associated Pontryagin flow equations are
\begin{align}
\dot{\hat U}(t)
&=
\frac{\partial \mathcal H_P}{\partial \hat P^\dagger}
=
-i\hat H(\mathbf u(t))\hat U(t),
\\
\dot{\hat P}(t)
&=
-\frac{\partial \mathcal H_P}{\partial \hat U^\dagger}= 
-i\hat H(\mathbf u(t))\hat P(t)
+
\frac{\partial f_0}{\partial \hat U^\dagger} \, .
\end{align}
The first equation simply recovers the Schr\"odinger equation for the evolution operator, while the second equation determines the backward propagation of the adjoint operator. Since the running cost $f_0(t)$ depends explicitly on $\hat U(t)$, the adjoint equation contains an inhomogeneous driving term proportional to $\partial f_0/\partial \hat U^\dagger$, which penalizes transient population of the excited state. In practice, we perform the quantum propagation in the rotating-frame picture, which converts the phase factor $e^{i \phi_+(t)}$ of Eq.~\eqref{eq:Hu} into an effective detuning $\dot{\phi}_+(t)$ on the $\ket{0}$ state. Precisely, by setting $\ket{\psi(t)} = \hat R(t) \ket{\widetilde{\psi}(t)}$ with $\hat R(t)=\exp[-i\phi_+(t)\ket{0}\bra{0}]$, we obtain in the rotating frame
{
\begin{eqnarray}
&& \widetilde{\hat{H}}(\mathbf{u}(t)) =  \left[ \frac 1 2 \Omega_{+}(t)  \ket{e}\bra{0} + \frac 1 2 \Omega_{\pi}(t) \ket{e}\bra{1}  \right.\nonumber \\
& + & \left. \frac 1 2 \Omega_{-}(t) \ket{e}\bra{2} + \mathrm{H.c.} \right] - \dot{\phi}_+(t) \ket{0} \bra{0}.
\end{eqnarray}
}

The adjoint operator satisfies the terminal condition fixed by the fidelity gradient,
\begin{equation}
\label{eq:adjointTerminal}
\hat P(t_f)
=
-\frac{\partial \Phi}{\partial \hat U^\dagger(t_f)}
=
\frac{1}{16}\,
\mathrm{Tr}\!\left[\hat U_\mathrm{target,4}^{\dagger}\hat U(t_f)\right]\,
\hat U_\mathrm{target,4}.
\end{equation}

For the unconstrained control problem considered here, the optimal controls $\mathbf{u}^*(t)$ follow from the stationarity condition
\begin{equation}
\frac{\partial \mathcal{H}_P}{\partial \mathbf{u}}=0
\end{equation}
along the optimal trajectories. The unitary evolution $\hat{U}(t)$ is obtained by forward propagation of the Schr\"odinger equation, while the adjoint operator $\hat{P}(t)$ is obtained by backward propagation from the terminal condition~\eqref{eq:adjointTerminal}.  

\section{Hahn-echo sequence}\label{secA1}
The unitary operator of the Hahn echo acting on the initial state $\ket{2,\mathbf{p}}$, with general momentum $\mathbf{p}$,  reads
\begin{multline}
    \hat U_\mathrm{echo}(\phi_+,\mathbf{p})=\\\hat H_3^\dagger\hat P(\phi_+)\hat U_\mathrm{free}(\mathbf{p})\hat X_3\hat U_\mathrm{free}(\mathbf{p})\hat X_3\hat U_\mathrm{free}(\mathbf{p})\hat H_3,
\end{multline}
where $\hat H_3$ and $\hat X_3$ are the SU(3) Hadamard and generalized Pauli-$X$ gates, respectively, and $\hat P(\phi_+)$ is a phase gate accounting for the phase jump imprinted by the $\sigma_+$ laser during the dark time.
These unitary operators can be expressed in matrix form as
\begin{equation}
\begin{aligned}
\hat H_3 &= 
\frac{1}{\sqrt{3}}\begin{pmatrix}
1 & 1 & 1\\
1 & e^{-2\pi i/3} & e^{2\pi i/3}\\
1 & e^{2\pi i/3} & e^{-2\pi i/3}
\end{pmatrix},\\
\hat X_3 &= 
\begin{pmatrix}
0 & 0 & 1\\
1 & 0 & 0\\
0 & 1 & 0
\end{pmatrix},\\
\hat P(\phi_+) &= 
\begin{pmatrix}
e^{i\phi_+} & 0 & 0\\
0 & 1 & 0\\
0 & 0 & 1
\end{pmatrix}.
\end{aligned}
\end{equation}
Each free-evolution interval has duration $T/3$, where $T$ is the time interval between the initial and final Hadamard gates. The corresponding free-evolution operator $\hat U_\mathrm{free}$ can be written as
\begin{equation}
\begin{aligned}
\hat U_\mathrm{free}(\mathbf{p}) = 
\begin{pmatrix}
e^{-iE_0T/3\hbar} & 0 & 0\\
0 & e^{-iE_1T/3\hbar} & 0\\
0 & 0 & e^{-iE_2T/3\hbar}
\label{Ufree}
\end{pmatrix},
\end{aligned}
\end{equation}
where 
\begin{equation}
\begin{aligned}
    E_0&=\frac{(\mathbf{p}+2\hbar\kappa_r\mathbf{\hat y})^2}{2m}-4E_r=\frac{p^2}{2m}+\frac{4\hbar\kappa_r\mathbf{p}\cdot\mathbf{\hat y}}{2m},\\
    E_1&=\frac{(\mathbf{p}+\hbar\kappa_r(\mathbf{\hat x}+\mathbf{\hat y}))^2}{2m}-2E_r=\frac{p^2}{2m}+\frac{2\hbar\kappa_r\mathbf{p}\cdot(\mathbf{\hat x}+\mathbf{\hat y})}{2m},\\
    E_2&=\frac{p^2}{2m},
    \label{E0E1E02}
\end{aligned}   
\end{equation}
are the energies of the three bare states after correcting for the recoil energy shifts, so no extra phase shift occurs for atoms initially in $\ket{2,p=0}$. $E_r=\hbar^2\kappa_r^2/2m$ is the recoil energy. In practice, we correct these shifts by red-detuning from resonance the $\sigma_+$ and $\pi$ lasers by $4E_r$ and $2E_r$, respectively.  

Because the two $\hat X_3$ gates cyclically permute the three internal states, each interferometer path spends an equal duration $T/3$ in each state. The intermediate sequence between the two Hadamard gates, $\hat U=\hat U_\mathrm{free}\hat X_3\hat U_\mathrm{free}\hat X_3\hat U_\mathrm{free}$, therefore becomes
\begin{equation}
    \hat U = e^{-i(E_0+E_1+E_2)T/3\hbar}
\hat X^\dagger_3.
\end{equation}
Thus, the accumulated dynamical phase is common to the three interferometer arms and can be factored out as a global phase.

\section{Experimental details}\label{secA2}
A gas of $^{87}\mathrm{Sr}$ atoms is successively Doppler cooled on the broad $^1S_0 \rightarrow {}^1P_1$ transition and the intercombination line $^1S_0 \rightarrow {}^3P_1$, before being loaded into an optical dipole trap (ODT) \cite{yang2015high}. The gas, which is initially distributed equally among the ten Zeeman ground states, is partially polarized by optically pumping the population in the positive $m_F$ states to $m_F = 9/2$, define as state $\ket{2}$. The population in the negative $m_F$ states is retained to induce thermalization via $s$-wave collisions during evaporative cooling \cite{Hasan_2022}. The final temperature of the gas following evaporative cooling is $0.21(1)\,T_F$, where $T_F = 190(9)\,\mathrm{nK}$ is the Fermi temperature.

After evaporative cooling, the ODT is switched off and the atoms are interrogated by the tripod beams in free fall. At the time of interrogation, the vertical \textit{rms} width of the cloud is measured to be $8(1)\,\mu\mathrm{m}$ from time-of-flight expansion measurements. The \textit{rms} widths along the two principal axes in the horizontal plane are estimated to be $9(1)\,\mu \mathrm{m}$ and $16(2)\,\mu \mathrm{m}$ from the temperature of the gas, the final powers of the ODT beams at the end of the evaporative cooling and the angle between them. The tripod beams ($1/e^2$ waist $\sim1\,\mathrm{mm}$) couple three internal states to a common excited state, as shown in Fig.~\ref{fig:2}(a), with their frequencies tuned to recoil-compensated resonances using acousto-optic modulators. This tripod scheme is isolated from the neighbouring excited states by approximately $12\Omega_\mathrm{max}$ at the experimental bias field of $ 67.0(3)\,\mathrm{G}$. Here, $\Omega_\mathrm{max}/(2\pi)=0.62(2)\,$MHz is the maximum Rabi frequency induced in each individual transition during gate operations. The change in the atomic internal state is associated with the absorption of a photon from one laser and its subsequent stimulated emission into another, which entangles the atomic spin with the momentum acquired from photon recoils. Since the momentum dispersion of the ultracold gas is $0.293(10)\hbar\kappa_r$,
this enables the measurement of the tripod ground-state populations from the momentum distribution in the far-field regime via fluorescence imaging on the $^1S_0 \rightarrow {}^1P_1$ transition. This far-field regime is accessed by a long time-of-flight expansion of the gas for about $15~\mathrm{ms}$. 

The negative $m_F$ states do not participate in the dynamics, since the tripod beams are detuned by approximately $40\Omega_\mathrm{max}$ from the nearest transition involving these states. Still, they contribute a fluorescence background at $(0,0)\hbar\kappa_r$, overlapping with the signal from state $\ket{2}$. This background is independently measured by transferring the population in $\ket{2}$ to $\ket{0}$ using stimulated Raman adiabatic passage. The residual signal at $(0,0)\hbar\kappa_r$, corresponding to negative states, is subsequently subtracted from all measured data.

\section{Generation of experimental pulses}\label{secA3}
The OCT pulses (Fig.~\ref{fig:3}(a)~and~(b)) are generated using a combination of acousto-optic and electro-optic modulators. The amplitude modulation (AM) envelope is produced by a programmable dual-channel arbitrary waveform generator (Rigol DG4102) with a temporal resolution of 10~ns. This envelope is mixed with an RF carrier using a mixer (Mini-Circuits ZX05-1-S+), where the carrier frequency is tuned to the relevant optical transition. The resulting AM RF signal is amplified and applied to the acousto-optic modulator (AOM). The modulation bandwidth is mainly limited by the rise time of the AOM ($\sim0.2\,\mu s$).

To calibrate the response of the circuit, we apply a linear input ramp and we record the diffracted power from the AOM, which is linked to the Rabi frequency via independent Rabi oscillation measurements. The required input waveform for a target Rabi amplitude profile is obtained through an inverse mapping between the measured Rabi frequency and the applied input signal. Since the spectral width of the AM envelope ($\sim 150~\mathrm{kHz}$) is well below the modulation bandwidth, the calibration remains valid over the entire pulse duration.

The phase modulation on the $\sigma_+$ beam is implemented using an electro-optic modulator (EOM). The temporal profile of the phase ramp is generated by modulating the input voltage to a high-gain amplifier (Thorlabs HVA200) of sufficient slew rate ($400\,\textrm{V}/\mu$s) driving the EOM (halfwave voltage $\sim150\,$V). The resulting phase modulation, shown in Fig.~\ref{fig:3}(a)~and~(b), is measured via homodyne detection by interfering the modulated beam with an unmodulated reference beam at the same optical frequency.

\section{Quantum process tomography}\label{secA4}
The experimental gates are benchmarked via a quantum process tomography. The action of the quantum channel $\hat \varepsilon$, defined in Eq.~\eqref{eqn:8}, on the input system $\hat{\rho}\in\mathcal{L}(\mathscr{H}_\mathrm{in})$, can be written in the operator-sum representation as
\begin{equation} {\hat{\varepsilon}}\left({\hat{\rho}}\right)=\sum_{i}{\hat{K}}_i{\hat{\rho}}{\hat{K}}_i^\dagger, 
\end{equation}
where $\hat{K}_i$ are the Kraus operators, satisfying $\sum \hat{K}_i^\dagger \hat{K}_i=\hat I$, which characterize the quantum channel acting on the system~\cite{Nielsen_Chuang_2010}. These operators can be written in the normalized Gell-Mann operator basis $\{\hat{\lambda}_m\}$ (defined in Table~\ref{tab:gm_basis}) as $\hat{K}_i = \sum_{j
} c_{i j}\hat \lambda_j$, where $\mathrm{Tr}\,[\hat \lambda_m\hat\lambda_n]=\delta_{mn}$. Substituting this into the above expression yields
\begin{equation}
{\hat{\varepsilon}}\left({\hat{\rho}}\right)=\sum_{m,n}{\hat{\lambda}}_m{\hat{\rho}}{\hat{\lambda}}_n^\dagger\chi_{mn},
\end{equation}
where $\chi_{mn}=\sum_i c_{im} c_{in}^*$ is the process matrix, constructed from the expansion coefficients of the Kraus operators in the Gell-Mann operator basis. 

\begin{table}[t]
\caption{\label{tab:gm_basis}%
Normalized Gell--Mann operator basis used to represent the Kraus operators.}
\begin{ruledtabular}
\begin{tabular}{c c | c c}
Operator &\multicolumn{1}{c}{Matrix} &  &  \\[.2em]
\cmidrule{1-2}
$\hat{\lambda}_0$ & $\dfrac{1}{\sqrt{3}}
\begin{pmatrix}
1&0&0\\
0&1&0\\
0&0&1
\end{pmatrix}$ & $\hat{\lambda}_5$ & $\dfrac{1}{\sqrt{2}}
\begin{pmatrix}
0&0&-i\\
0&0&0\\
i&0&0
\end{pmatrix}$ \\[1.6em]
$\hat{\lambda}_1$ & $\dfrac{1}{\sqrt{2}}
\begin{pmatrix}
0&1&0\\
1&0&0\\
0&0&0
\end{pmatrix}$ & $\hat{\lambda}_6$ & $\dfrac{1}{\sqrt{2}}
\begin{pmatrix}
0&0&0\\
0&0&1\\
0&1&0
\end{pmatrix}$ \\[1.6em]
$\hat{\lambda}_2$ & $\dfrac{1}{\sqrt{2}}
\begin{pmatrix}
0&-i&0\\
i&0&0\\
0&0&0
\end{pmatrix}$ & $\hat{\lambda}_7$ & $\dfrac{1}{\sqrt{2}}
\begin{pmatrix}
0&0&0\\
0&0&-i\\
0&i&0
\end{pmatrix}$ \\[1.6em]
$\hat{\lambda}_3$ & $\dfrac{1}{\sqrt{2}}
\begin{pmatrix}
1&0&0\\
0&-1&0\\
0&0&0
\end{pmatrix}$ & $\hat{\lambda}_8$ & $\dfrac{1}{\sqrt{6}}
\begin{pmatrix}
1&0&0\\
0&1&0\\
0&0&-2
\end{pmatrix}$ \\[1.6em]
$\hat{\lambda}_4$ & $\dfrac{1}{\sqrt{2}}
\begin{pmatrix}
0&0&1\\
0&0&0\\
1&0&0
\end{pmatrix}$ & & \\[.8em]
\end{tabular}
\end{ruledtabular}
\end{table}












To experimentally reconstruct the ${\chi}$ matrix, the atoms are prepared in an informationally complete set of input states $\{\hat \rho_k=\ket{\psi_k}\bra{\psi_k}\}$, where $\ket{\psi_k}$ are defined as
\begin{equation}\label{input_states}
\begin{aligned}
&\ket{0}, \, \ket{1}, \, \ket{2}, \\
&\frac{1}{\sqrt{2}}(-\ket{0}+\ket{1}), \, \frac{1}{\sqrt{2}}(-i\ket{0}+\ket{1}), \, \frac{1}{\sqrt{2}}(\ket{1}+\ket{2}), \\
&\frac{1}{\sqrt{2}}(-i\ket{1}+\ket{2}), \, \frac{1}{\sqrt{2}}(\ket{0}+\ket{2}),\,\textrm{and}\,\frac{1}{\sqrt{2}}(-i\ket{0}+\ket{2}).
\end{aligned}
\end{equation}

For this set of input states, $\ket{0}$ and $\ket{1}$ are prepared from $\ket{2}$ via STIRAP, while the coherent superposition states are prepared using pairwise two-photon Raman rotations. The experimental outcomes for these input states are measured for a set of projection operators $\mathcal{\hat P}_{ij}=\hat R_i^\dagger\ket{j}\bra{j}\hat R_i$, where $\ket{j}\bra{j}=\{\ket{0}\bra{0},\ket{1}\bra{1},\ket{2}\bra{2}\}$ and {$\hat R_i$} are the pairwise Raman rotations. These rotation operations are defined as
\begin{equation}\label{Raman_rotations}
\begin{array}{cc}
\hat I=
\begin{pmatrix}
1 & 0 & 0\\
0 & 1 & 0\\
0 & 0 & 1
\end{pmatrix},
&
\hat R_1=\dfrac{1}{\sqrt{2}}
\begin{pmatrix}
1 & -1 & 0\\
1 & 1 & 0\\
0 & 0 & \sqrt{2}
\end{pmatrix},
\\[1.6em]

\hat R_2=\dfrac{1}{\sqrt{2}}
\begin{pmatrix}
1 & -i & 0\\
-i & 1 & 0\\
0 & 0 & \sqrt{2}
\end{pmatrix},
&
\hat R_3=\dfrac{1}{\sqrt{2}}
\begin{pmatrix}
\sqrt{2} & 0 & 0\\
0 & 1 & 1\\
0 & -1 & 1
\end{pmatrix},
\\[1.6em]

\hat R_4=\dfrac{1}{\sqrt{2}}
\begin{pmatrix}
\sqrt{2} & 0 & 0\\
0 & 1 & -i\\
0 & -i & 1
\end{pmatrix},
&
\hat R_5=\dfrac{1}{\sqrt{2}}
\begin{pmatrix}
1 & 0 & 1\\
0 & \sqrt{2} & 0\\
-1 & 0 & 1
\end{pmatrix},
\\[1.6em]

\mathrm{and}\, \hat R_6=\dfrac{1}{\sqrt{2}}
\begin{pmatrix}
1 & 0 & -i\\
0 & \sqrt{2} & 0\\
-i & 0 & 1
\end{pmatrix}.
&
\end{array}
\end{equation}




The expectation value of the projection operator $\mathcal{\hat P}_{ij}$ is
\begin{equation}\label{eqn:20}
\begin{aligned}
\Braket{\mathcal{P}_{ijk}} & = \mathrm{Tr}\!\left[\mathcal{\hat P}_{ij} \, \hat\varepsilon(\hat\rho_k)\right]\\
&= \, \mathrm{Tr}\!\left[\ket{j}\bra{j}\hat R_i \hat \varepsilon(\hat \rho_k)\hat R_i^\dagger\right]\\
&= \sum_{m,n} \chi_{mn} \, B_{mnijk},
\end{aligned}
\end{equation}
where
\begin{equation}
B_{mnijk} = \mathrm{Tr}\!\left[\ket{j}\bra{j}\hat R_i \hat \lambda_m \hat \rho_k \hat \lambda_n^\dagger \hat R_i^\dagger\right].
\end{equation}
If $\mathscr{M}_{ijk}$ are the measured outcomes for the projection $\mathcal{\hat P}_{ij}$ when the quantum process acts on input states, then the unknown components of ${\chi}$ can be obtained by solving the following optimization problem:
\begin{equation}\label{eqn:22}
\min_{\chi} \;\; \left\|\, \mathscr{M}_{ijk} - \Braket{\mathcal{P}_{ijk}} \right\|_2,
\end{equation}

Since a physical quantum channel must be completely positive and trace preserving, its process matrix $\chi$ reconstructed from experimental data is obtained by solving Eq.~\eqref{eqn:22} subject to the following constraints:
\begin{equation}
\chi \succeq 0, \qquad \sum_{m,n}\chi_{mn}{\hat{\lambda}}_n^\dagger{\hat{\lambda}}_m=I
\end{equation}
which is implemented using a convex optimization method.


The Choi matrix (defined in Eq.~\eqref{eqn:9}) can be obtained from the $\chi$ matrix through the transformation
\begin{equation}
\mathcal{C}_\varepsilon
=
\frac{1}{d}
\sum_{m,n}
\chi_{mn}
\ket{\!\,\lambda_m}\!\rangle
\langle\!\bra{\lambda_n},
\end{equation}
where
\begin{equation}
\ket{\!\,\lambda_m}\!\rangle
\equiv
\sum_{k=0}^{d-1}
\ket{k}\otimes \hat{\lambda}_m \ket{k}.
\end{equation}

\section{Effect of finite momentum dispersion of atoms}\label{secAxx}

\begin{figure*}[t!]
\centering
\includegraphics[width=0.9\textwidth]{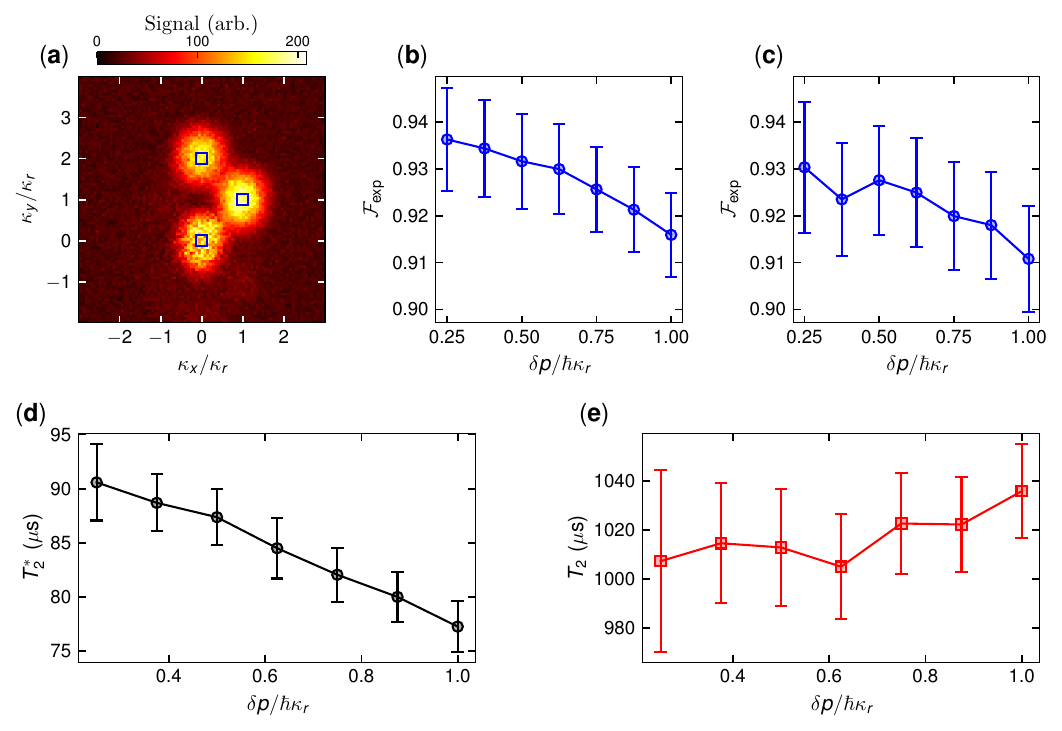}
\caption{\textbf{(a)} Momentum distribution of the atomic gas after the application of the Hadamard gate to the state $\ket{2}$, measured using fluorescence imaging after a 15 ms time-of-flight expansion. This distribution is obtained after subtracting the fluorescence background from negative states centered at $(0,0)\hbar\kappa_r$. For the QPT, the population is integrated within the momentum range of $\hbar\kappa_r[-0.125,0.125]$ centered on each momentum peak, represented by the blue boxes. \textbf{(b)}--\textbf{(c)} Measured gate fidelities for the $\hat H_3$ and $\hat X_3$ operations as a function of momentum range size $\delta p$. \textbf{(d)} Dependence of the Ramsey dephasing time $T_2^*$ on momentum sampling size. The decrease in $T_2^*$ for larger windows results from averaging over spatial interference fringes shown in Fig.~\ref{fig:2}(e). This inhomogeneous dephasing is corrected by the generalized Hahn-echo sequence, leading to a homogeneous spatial distribution of the bare state populations (Fig.~\ref{fig:2}(f)). Consequently, the $T_2$ homogeneous dephasing time remains largely insensitive to the momentum sampling size, as shown in \textbf{(e)}.}
\label{fig:a1}
\end{figure*}	

To compensate for the recoil-energy shifts of the states $\ket{0}$ and $\ket{1}$, the $\sigma_+$ and $\pi$ beams are red-detuned from their bare resonances by $4E_r$ and $2E_r$, respectively. Hence, the three tripod bare states are degenerated for atoms with $p=0$. However, atoms with nonzero momentum acquire an additional dynamical phase during the experimental sequence, consistent with Eqs.~\eqref{Ufree} and \eqref{E0E1E02}. The resulting dephasing affects the full QPT sequence. The latter consists of the Hahn-echo gates together with the input and output Raman rotations, each with a pulse duration of $\sim6~\mu$s. In addition, the experimental control system imposes a minimum delay of $2~\mu\mathrm{s}$ between successive operations, resulting in a total QPT duration of $26~\mu$s. The finite momentum dispersion also reduces the Raman-pulse fidelity due to uncontrolled residual two-photon detuning. The measured momentum width of $0.293(10)\hbar\kappa_r$ corresponds to a detuning of $\sim0.15\Omega_\mathrm{Raman}$, where $\Omega_{\mathrm{Raman}}/2\pi\sim40~\mathrm{kHz}$ is the two-photon Raman Rabi frequency.

To reduce the contribution of these momentum-dependent effects to the tomography sequence, the population in each state is integrate within a narrower momentum range around the center of each peak of the time-of-flight distribution. We use a momentum range size of $\delta p=0.25\hbar\kappa_r$, as depicted by blue boxes in Fig.~\ref{fig:a1}(a). Nevertheless, Fig.~\ref{fig:a1}(b)~and~(c) shows the measured gate fidelities (using Eq.~\eqref{eqn:11}) as a function of the momentum range size for $\hat H_3$ and $\hat X_3$. Thus, the reduction in fidelity for larger sampling windows can be attributed to inhomogeneous Doppler dephasing due to the finite momentum distribution of the gas. 

This post-selection of a finite momentum range also affects the measured $T_2^*$ time in the Ramsey interferometer, which decreases monotonically with increasing window size (Fig.~\ref{fig:a1}(d)). In contrast, the momentum dispersion has no clear impact on the Hahn-echo dephasing time $T_2$ as expected for a homogeneous dephasing time, as shown in Fig.~\ref{fig:a1}(e). This confirms the role of the momentum dispersion as the main inhomogeneous dephasing mechanism. 

\section{Simulations with effective Hamiltonian}\label{secA_simulation}
\subsection{Model}
We perform numerical simulations using a rotating-wave-approximation effective Hamiltonian $\hat H_\mathrm{eff}$, including the complete Zeeman manifolds of the $^1S_0, F=9/2$ ground and the $^3P_1, F=9/2$ excited levels. The energy separations between substates are calculated from the Zeeman splitting induced by the $67\,\mathrm{G}$ bias magnetic field. All allowed transitions are driven by the three tripod beams with polarizations $\sigma_\pm$ and $\pi$, that are supposed to be perfect. We take into account the spontaneous emission by inserting a non-Hermitian term $-i\Gamma/2$ on the diagonal terms of $\hat H_\mathrm{eff}$ associated with excited states. $\Gamma/2\pi=7.5\,\mathrm{kHz}$ is the linewidth of the excited states. We solve the Schrödinger equation by considering $\hat H_\mathrm{eff}$ piecewise constant with time steps $\delta t<0.25\,\mathrm{ns}$ (sufficiently small for the fidelity to converge). Here, we apply the evolution operators $\hat U_n = \mathrm{e}^{-i\hat H_\mathrm{eff}(n\delta t)\delta t/\hbar}$ successively to the state vector. 
The use of a non-Hermitian effective Hamiltonian rather than a full Lindblad description is justified by the fact that losses remain typically below $5\%$ for all simulations, and spin-orbit coupling filters out, after time-of-flight, the atoms that have undergone spontaneous emission.

\subsection{Numerical simulation of SPAM operations and QPT}

First, we use the above model to find the optimal experimental parameters for the Raman rotations that minimize the error during SPAM in the QPT sequence (Eq.~\ref{Raman_rotations}). 
The presence of excited states outside the tripod scheme creates an imbalance in the light shifts, resulting in unwanted two-photon detunings. This imbalance is compensated for by adjusting the Rabi frequencies of each beam. The optimum parameters correspond to the best compromise between the reduction of the tripod excited state population and the strength of neighboring transitions that can cause Raman coupling outside the qutrit manifold.
These Raman parameters are listed in Table~\ref{tab:Raman_param} for each SPAM rotation operation. We evaluate the SPAM optimization procedure by computing the fidelity of the identity gate using the QPT sequence. For an atom at rest ($\delta p=0$), the optimized parameters yield a fidelity of $\sim0.98$. For experimental implementation, we use the same pulse generation setup as Appendix~\ref{secA3}, with the phase shift $\phi_-$ on the $\sigma_-$ beam imprinted by an additional EOM.

\begin{table}[h]
\caption{\label{tab:Raman_param}%
Raman parameters used for SPAM. Here, $\Omega_{\pm,\pi}$ denote the tripod Rabi frequencies and $\Delta$ the corresponding single-photon detunings found using the numerical model discussed in the main text. The duration of the square Raman pulses is $\sim6\,\mu$s. The Rabi frequencies and detunings have units of MHz.}
\begin{ruledtabular}
\begin{tabular}{c c c c c c c}
Rotation & $\Omega_+/2\pi$ & $\Omega_\pi/2\pi$ & $\Omega_-/2\pi$ & $\Delta/2\pi$& $\phi_+$ & $\phi_-$  \\
\midrule
$\hat R_1$ & 0.315 & 0.345 & -- & 1.256&$\pi/2$ & 0 \\
$\hat R_2$ & 0.315 & 0.345 & -- & 1.256&0 & 0  \\ 
$\hat R_3$ & -- & 0.284 & 0.380 & 1.250&0 & $\pi/2$ \\
$\hat R_4$ & -- & 0.284 & 0.380 & 1.250&0 & 0 \\
$\hat R_5$ & 0.310 & -- & 0.349 & 1.253&0 & $\pi/2$ \\
$\hat R_6$ & 0.310 & -- & 0.349 & 1.253&0 & 0 \\
\end{tabular}
\end{ruledtabular}
\end{table}

Secondly, we utilize the model to extract the simulated values of the fidelities presented in Table~\ref{tab:gate_fidelities}, following the tomography procedure presented in Appendix~\ref{secA4}. $\mathcal{F}_\mathrm{sim}$ is obtained by simulating the effect of the chosen gate on the input states given by Eq.~\eqref{input_states}, using the experimentally recorded control curves of Fig.~\ref{fig:3}(a) and (b), before numerically applying the pairwise Raman rotations of Eq.~\eqref{Raman_rotations} to the propagated state. The resulting state is projected onto the qutrit subspace and normalized before reconstructing the corresponding Choi matrix, which is inserted into Eq.~\eqref{eqn:11} to obtain the gate fidelity. Next, we include experimental initial momentum and position distributions in the simulations, leading to the fidelity $\mathcal{F}_\mathrm{sim}^{\delta p,\delta r}$. Finally, $\mathcal{F}_\mathrm{sim}^{\delta p,\delta r + \mathrm{SPAM}}$ is computed with SPAM Raman rotations using the optimum parameters from Table~\ref{tab:Raman_param}.

The simulations are performed with $N=10\,000$ atoms. The mean values and standard deviations for $\mathcal{F}_\mathrm{sim}^{\delta p,\delta r}$ and $\mathcal{F}_\mathrm{sim}^{\delta p,\delta r + \mathrm{SPAM}}$ are obtained with a $15$~ms time of flight, an integration over the momentum range $\hbar\kappa_r[-0.125, 0.125]$, and a statistical analysis over 200 realizations where the position of the atoms is resampled from the initial distribution (without resampling the momenta).

\section{Error budget}\label{secA5}

To estimate the contributions of the various experimental imperfections to the gate infidelity, we numerically simulate the gates and the QPT sequence, while successively incorporating the various known experimental error sources, as explained in Appendix~\ref{secA_simulation}. 

For the ideal control pulses obtained from OCT, the intrinsic gate infidelity is approximately $1\times10^{-3}$. Next, we simulate the gate using the experimental control pulses shown in Fig.~\ref{fig:3}(a)~and~(b)). This increases the simulated infidelity by $\sim 9\times10^{-3}$. When accounting for the initial horizontal size of the atomic cloud and the initial momentum distribution of the atoms, and integrating within the range $\hbar\kappa_r[-0.125, 0.125]$ after time-of-flight, the infidelity increases by $\sim 1\times10^{-2}$. Finally, including the finite-duration Raman pulses used in the QPT sequence increases the infidelity to approximately $4.5\times10^{-2}$. This indicates that SPAM errors during the QPT sequence provide the dominant contribution to the observed gate infidelity. 

The remaining error sources are treated independently and their contribution simply added to the error budget. The shot-to-shot magnetic field fluctuations are estimated to contribute an infidelity of approximately $1.2\times10^{-2}$. The laser intensity fluctuation is modeled as white noise and is found to contribute to the infidelity by less than $1\times10^{-3}$. We finally include the contribution of laser phase noise from the experimental laser noise spectrum. The resulting infidelity is less than $1\times10^{-3}$. We note that the laser noises, phase and amplitude, are negligible at the level of the experimental uncertainty.

The infidelity error budget is given in Table~\ref{tab:error_budget}.
\begin{table}[H]
\caption{\label{tab:error_budget}%
Contributions to the gate infidelity. The first row gives the intrinsic gate infidelity of the ideal OCT pulses. The subsequent three rows show the cumulative infidelity obtained after incorporating the experimentally realized pulse profiles, momentum dispersion, and the Raman operations used in the QPT sequence. The remaining error sources are evaluated independently.}
\begin{ruledtabular}
\begin{tabular}{l c}
Error source & Infidelity ($\times10^{-3}$) \\
\midrule
\multicolumn{2}{l}{\textit{Intrinsic gate limit}} \\
Ideal OCT pulses & 1 \\
\midrule
\multicolumn{2}{l}{\textit{Cumulative experimental model}} \\
Experimental pulses & 10 \\
+ $\delta p$,$\delta r$ & 20 \\
+ QPT Raman pulses (SPAM) & 45 \\
\midrule
\multicolumn{2}{l}{\textit{Additional independent contributions}} \\
Magnetic-field fluctuations & 12 \\
Laser-intensity noise & $<1$ \\
High-frequency laser phase noise & $<1$ \\
\end{tabular}
\end{ruledtabular}
\end{table}

\bibliography{sn-bibliography}

\end{document}